\documentstyle[12pt,epsfig]{article}
\def\lsim{\raise0.3ex\hbox{$<$\kern-0.75em\raise-1.1ex\hbox{$\sim$}}}
\def\gsim{\raise0.3ex\hbox{$>$\kern-0.75em\raise-1.1ex\hbox{$\sim$}}}

\def\pom{{I\!\!P}}

\def\beq{\begin{equation}}
\def\eeq{\end{equation}}
\def\bea{\begin{eqnarray}}
\def\eea{\end{eqnarray}}
\def\bq{\begin{quote}}
\def\eq{\end{quote}}

\newcommand{\rr}{\mbox{\boldmath $r$}}

\def\gappeq{\mathrel{\rlap {\raise.5ex\hbox{$>$}}
{\lower.5ex\hbox{$\sim$}}}}

\def\lappeq{\mathrel{\rlap{\raise.5ex\hbox{$<$}}
{\lower.5ex\hbox{$\sim$}}}}

\def\Toprel#1\over#2{\mathrel{\mathop{#2}\limits^{#1}}}

\newcommand{\rk}{\mbox{\boldmath $k$}}

\def\pom{{I\!\!P}}

\begin{document}
\pagestyle{empty}
\begin{center}
{\bf NUCLEAR EXCLUSIVE VECTOR MESON PHOTOPRODUCTION}
\\

\vspace*{1cm}
 V.P. Gon\c{c}alves $^{1}$, M.V.T. Machado  $^{2,\,3}$\\
\vspace{0.3cm}
{$^{1}$ Instituto de F\'{\i}sica e Matem\'atica,  Universidade
Federal de Pelotas\\
Caixa Postal 354, CEP 96010-090, Pelotas, RS, Brazil\\
$^{2}$ \rm Universidade Estadual do Rio Grande do Sul - UERGS\\
 Unidade de Bento Gon\c{c}alves. CEP 95700-000. Bento Gon\c{c}alves, RS, Brazil\\
$^{3}$ \rm High Energy Physics Phenomenology Group, GFPAE  IF-UFRGS \\
Caixa Postal 15051, CEP 91501-970, Porto Alegre, RS, Brazil}\\
\vspace*{1cm}
{\bf ABSTRACT}
\end{center}

\vspace*{1cm} \noindent

\vspace*{1.5cm} \noindent \rule[.1in]{17cm}{.002in}

\vspace{-3.5cm} \setcounter{page}{1} \pagestyle{plain} We
calculate the nuclear  cross section for vector meson exclusive
photoproduction within the QCD color dipole picture and in the
Regge approach. For the former approach, we have considered the
phenomenological saturation model, whereas  for the latter  we use
a model based on the dipole Pomeron framework. Theoretical
estimates for scattering on both  light and heavy nuclei are given over a large
range on  energy. \vspace{1.5cm}

\section{Introduction}

 Exclusive vector meson production by real and virtual photons is
 an outstanding process  providing important information on the transition
region from the soft dynamics (at low virtualities of the photon
$Q^2$) to the hard perturbative regime at high $Q^2$
\cite{predazzi,crittenden}. In principle, a perturbative approach
is only justified if a hard scale is present in the process, e.g. the
photon virtuality and/or a large mass of the vector meson. For photoproduction of light mesons,
 such scale is not present and one has to rely on non-perturbative models. In general, a simple Regge pole phenomenology, with a soft Pomeron having intercept larger than one, is enough to describe the energy dependence of the meson cross section at the present accelerators. On some pQCD approaches, as the saturation model, even this soft process can be described, where the transition is set by the saturation scale. Both models give an effective Pomeron intercept increasing with photon virtuality and meson mass. For  $\phi$ and $J/\Psi$ mesons,  the Pomeron intercept is considerably large and consistent with that one obtained in usual pQCD approaches. Despite the good agreement for the currently  available energies, an extrapolation to higher energies of the 
experimental fits implies a large growth for the cross section
which it would violate the unitarity at sufficiently high energies. Therefore,
dynamical modifications associated to the unitarity corrections
are also expected to be present in this process
\cite{mara,stasto}. Moreover, these effects should be enhanced in nuclear
processes \cite{glr,levin}. In particular, in the planned $eA$
colliders at HERA and RHIC, the experimental analyzes of the
exclusive vector meson production could be very useful to
constrain the QCD dynamics \cite{hera96,raju}.

Our goal in this paper is to investigate the high energy vector
meson exclusive  photoproduction on nuclei. In particular, we
improve the previous analyzes of vector meson production
\cite{klein} which are based on an extrapolation of the DESY-HERA experimental fits for the proton case. Here, we will consider unitarized cross sections {\it ab initio}. In order to do so,
 we consider two distinct and well established theoretical scenarios,
which do not violate the unitarity bound in the asymptotic regime to
be probed in future colliders. This allows us to analyze the
nuclear vector meson protoproduction as a potential process to
discriminate between these different theoretical approaches. First, we consider the color dipole
description of the $\gamma A \rightarrow VA$ ($V=\rho,\,\omega,\,
\phi,\,J/\Psi$) process, which is quite successful for the proton
case \cite{mara,stasto} and can be extended to nuclei targets via
Glauber-Gribov formalism. It is important to quote the pioneering papers \cite{Nikolaev:1992si,Benhar:df,Kopeliovich:1993gk,Kopeliovich:1993pw,Nemchik:1994fq}, where these issues were first addressed and their further developments \cite{Nikolaev:1999bq,Benhar:ue,Ivanov:2003gr} as well. Related calculations in the $k_{\perp}$-factorization approach (dipole approach is equivalent to it at leaging logarithmic aproximation) can be found in the Refs. \cite{igor1,igor2,igor3,igor4}. In the color dipole approach, the degrees of freedom are
the photon (color dipole) and meson wavefunctions as well as the
dipole-nuclei cross section. Such an approach enables to include
nuclear effects and parton saturation phenomenon. The latter one is
characterized by a typical momentum scale $Q_{\mathrm{sat}}$ (saturation
scale) and it has been constrained by
experimental results in deep inelastic scattering (DIS) and
diffractive DIS \cite{golecwus}. Here, we will use an extension of the phenomenological saturation model for  nuclear
targets \cite{armesto}. This model  reasonably describes  the
experimental data for the nuclear structure function  and has been
used to predict the nuclear inclusive and diffractive cross
sections for heavy quark photoproduction \cite{vicmagsat}. The nuclear saturation scale,
$Q_{\mathrm{s}\,A}$, provides the transition between the
color transparency  and the saturation regimes in the nuclear scattering. Concerning vector meson production, our starting point is
the recent work in Ref. \cite{sandapen}, where different meson
wavefunctions and dipole cross sections are considered for the
proton case. It is worth mentioning that although light meson
photoproduction to be a soft process by definition, it is
consistently described in the QCD color dipole picture whether
there is a suitable model for the soft-hard transition, as occurring in the 
the saturation model.

In order to compare the saturation approach  with a successful
nonperturbative formalism,  we consider a Regge inspired model
given by the dipole Pomeron framework \cite{dipolepom}. The reason for this
particular choice is that in this model the soft Pomeron having intercept equal one,
thus it does not violate unitarity for hadron-hadron and vector
meson production at higher energies. Moreover, it describes with
good agreement hadronic cross section and even DIS data in a wide
range of photon virtualities \cite{dippomdis}.  For meson production on proton
target, our starting point is the recent work of Ref.
\cite{martynov_predazzi}, where exclusive photoproduction by real
and virtual photons is described with good agreement. The
extension to nuclei is provided by the assumption of vector meson dominance and
the Glauber-Gribov formalism.

This paper is organized as  follows. In the next section, we
present a brief review of the exclusive meson production in the color
dipole picture for proton target and its extension to the nuclear
case. For the dipole-proton (nucleus) cross section we have considered the
phenomenological saturation model, which is shortly described.  In Sec. 3, the dipole Pomeron parameterization for  vector meson
photoproduction  is presented and its extension to nuclear targets is
considered. The results coming out from both models are presented and
discussed in Sec. 4.  Finally, in Section \ref{conc} we summarize our
conclusions.

\section{Vector meson production in the color dipole approach}
\label{sec:dipolePic}

Let us introduce the main formulas concerning the vector meson
production in the color dipole picture. First, we consider the
scattering process $\gamma p \rightarrow Vp$, where $V$ stands for
both light and heavy mesons. Further, one extends this approach to
 the nuclear case. The scattering process can be seen
in the target rest frame as a succession in time of three
factorizable subprocesses: i) the photon fluctuates in a
quark-antiquark pair (the dipole), ii) this color dipole interacts with the
target and, iii) the pair converts into vector meson final state.
Using as kinematic variables the $\gamma^* N$ c.m.s. energy
squared $s=W_{\gamma N}^2=(p+q)^2$, where $p$ and $q$ are the target and the
photon momenta, respectively, the photon virtuality squared
$Q^2=-q^2$ and the Bjorken variable $x=Q^2/(W_{\gamma N}^2+Q^2)$, the
corresponding imaginary part of the amplitude at zero momentum
transfer reads as \cite{nik},
\begin{eqnarray}
{\cal I}m \, {\cal A}\, (\gamma p \rightarrow Vp)  = \sum_{h, \bar{h}}
\int dz\, d^2\rr \,\Psi^\gamma_{h, \bar{h}}(z,\,\rr,\,Q^2)\,\sigma_{dip}^{\mathrm{target}}(\tilde{x},\rr) \, \Psi^{V*}_{h, \bar{h}}(z,\,\rr) \, ,
\label{sigmatot}
\end{eqnarray}
where $\Psi^{\gamma}_{h, \bar{h}}(z,\,\rr)$ and $\Psi^{V}_{h,
  \bar{h}}(z,\,\rr)$  are the light-cone wavefunctions  of the photon
  and vector meson, respectively. The quark and antiquark helicities are labeled by $h$ and $\bar{h}$
  and reference to the meson and photon helicities are implicitly understood. The variable $\rr$ defines the relative transverse
separation of the pair (dipole) and $z$ $(1-z)$ is the
longitudinal momentum fractions of the quark (antiquark). The basic
blocks are the photon wavefunction, $\Psi^{\gamma}$, the  meson
wavefunction, $\Psi_{T,\,L}^{V}$,  and the dipole-target  cross
section, $\sigma_{dip}^{\mathrm{target}}$.

In the dipole formalism, the light-cone
 wavefunctions $\Psi_{h,\bar{h}}(z,\,\rr)$ in the mixed
 representation $(r,z)$ are obtained through two dimensional Fourier
 transform of the momentum space light-cone wavefunctions
 $\Psi_{h,\bar{h}}(z,\,\rk)$ (see more details, e.g. in Refs. \cite{predazzi,stasto,sandapen}). The
 normalized  light-cone wavefunctions for longitudinally ($L$) and
 transversely ($T$) polarized photons are given by \cite{dgkp:97}:
\begin{eqnarray}
\Psi^{L}_{h,\bar{h}}(z,\,\rr)& = & \sqrt{\frac{N_{c}}{4\pi}}\,\delta_{h,-\bar{h}}\,e\,e_{f}\,2 z(1-z)\, Q \frac{K_{0}(\varepsilon r)}{2\pi}\,,
\label{wfL}\\
\Psi^{T(\gamma=\pm)}_{h,\bar{h}}(z,\,\rr) & = & \pm
\sqrt{\frac{N_{c}}{2\pi}} \,e\,e_{f}
 \left[i e^{ \pm i\theta_{r}} (z \delta_{h\pm,\bar{h}\mp} -
(1-z) \delta_{h\mp,\bar{h}\pm}) \partial_{r}
+  m_{f} \,\delta_{h\pm,\bar{h}\pm} \right]\frac{K_{0}(\varepsilon r)}{2\pi}\,,
\label{wfT}
\end{eqnarray}
where $\varepsilon^{2} = z(1-z)Q^{2} + m_{f}^{2}$. The quark mass
$m_f$ plays a role of a regulator when the photoproduction
regime is reached.  Namely, it prevents non-zero argument for the
modified Bessel functions $K_{0,1}(\varepsilon r)$ towards $Q^2\rightarrow 0$.
 The electric charge of the quark of flavor $f$ is given by $e\,e_f$.

For vector mesons, the light-cone wavefunctions are not known
in a systematic way and they are thus obtained through models. The
simplest approach  assumes a same vector  current as in the photon
case, but introducing  an additional vertex factor. Moreover, in
general a same functional form for the scalar part of the meson
light-cone wavefunction is chosen. Here, we follows the
analytically simple DGKP approach \cite{dgkp:97}, which is found to describe in
good agreement vector meson production as pointed out in Ref.
\cite{sandapen}. In this particular approach, one assumes
that the dependencies on $\rr$ and $z$ of the wavefunction are
factorised, with a Gaussian dependence on $\rr$. The DGKP
longitudinal and transverse meson light-cone wavefunctions are
given by \cite{dgkp:97},
\begin{eqnarray}
\Psi_{h,\bar{h}}^{V,L}(z,\,\rr) & = &
z(1-z)\, \delta_{h,-\bar{h}}\,\frac{\sqrt{\pi} f_V}
{2\sqrt{N_{c}}\,\hat{e}_{f}}\,f_{L}(z)\,\exp
\left[\frac{-\,\omega_{L}^{2}\, \rr^{2}}{2} \right] \;,
\label{dgkp_L}\\
\Psi_{h,\bar{h}}^{V,T(\gamma = \pm)}(z,\,\rr) &=& \pm
\left(\frac{i\omega_T^{2}\,r e^{\pm i\theta_{r}}}{m_{V}}\,
[z\delta_{h\pm,\bar{h}\mp} -
(1-z)\delta_{h\mp,\bar{h}\pm}] + \frac{m_{f}}{m_{V}}\,\delta_{h\pm,\bar{h}\pm}
\right) \nonumber \\ & & \hspace*{4cm} \times
\frac{\sqrt{\pi} f_V}{\sqrt{2 N_{c}}
\,\hat{e}_{f}}f_{T}(z)\,\exp
\left[\frac{-\omega_{L}^{2} \rr^{2}}{2} \right] \;.
\label{dgkp_T}
\end{eqnarray}
where $\hat{e}_f$ is the effective charge arising from the sum
over quark flavors in the meson of mass $m_V$. The following
values $\hat{e}_f = 1/\sqrt{2},\,1/3\sqrt{2},\, 1/3$ and $2/3$
stand for the $\rho$, $\omega$, $\phi$ and $J/\Psi$, respectively.
The coupling of the meson to electromagnetic current is labeled
by $f_V^2=3\,m_V\Gamma_{e^+e^-}/4\,\pi\alpha_{em}^2$ (see Table
\ref{tab:1}). The function $f_{T,\,L}(z)$ is given by the
Bauer-Stech-Wirbel model \cite{wsb:85}:
\begin{eqnarray}
f_{T,\,L}(z)= {\cal N}_{T,\,L}\,
\sqrt{z(1-z)}\,\exp \left[\frac{-\,m_{V}^{2}\,(z-1/2)^{2}}{2\,\omega^{2}_{T,\,L}}\right] \;.
\end{eqnarray}

The meson wavefunctions are constrained by the normalization
condition, which contains the hypothesis of a meson composed only of
quark-antiquark pairs,  and  by the electronic decay width
$\Gamma_{V\rightarrow e^+e^-}$. Both
conditions are respectively  given by \cite{bl:80,stasto},
\begin{eqnarray}
& & \sum_{h,\bar{h}}\int d^{2}\rr \, dz  \,
|\Psi^{V(\lambda)}_{h,\bar{h}}(z,\,\rr)|^{2} = 1\,,
\label{norm1}\\
& & \sum_{h,\bar{h}}\int \frac{d^{2}\rr}{(2 \pi)^2}\, \frac{dz}
{z(1-z)}\, [z(1-z) Q^2 + k^2+m_f^2]\, \Psi^{V}_{h,\bar{h}}(k,z)
\Psi^{\gamma *}_{h,\bar{h}}(k,z)= e f_V m_V\, (\varepsilon_{\gamma}^{*}\cdot\varepsilon_{V})\,.
\label{norm2}
\end{eqnarray}
The above constraints when used   on the DGKP wavefunction
produce the following relations \cite{sandapen},
\begin{eqnarray}
& & \omega_{L,\,T} =   \frac{ \pi f_{V}}{\sqrt{2\, N_c} \hat{e}_{f}}
\,\sqrt{I_{L,\,T}}\,,
\label{dgkpnorm1}\\
& & \int_{0}^{1} dz \; z(1-z)\,f_{L}(z) = \int_{0}^{1} dz \,\frac{2\,[z^{2} +
    (1-z)^{2}]\,\omega_{T}^{2} + m_{f}^{2}}{2\,m_{V}^{2}\,z(1-z)}\,f_{T}(z) =1
\,,
\label{dgkpnorm2}
\end{eqnarray}
where
\begin{eqnarray}
I_{L} & = & \int_{0}^{1} dz \,z^{2}(1-z)^{2} \,f_{L}^{2}(z) \,,\\
I_{T} & = &   \int_{0}^{1} dz \,\frac{[z^{2} + (1-z)^{2}]\,\omega_{T}^{2} + m_{f}^{2}}{m_{V}^{2}}\, f_{T}^{2}(z) \,.
\end{eqnarray}

The relations in Eq. (\ref{dgkpnorm1}) come from the
normalization condition, whereas the relations in Eq.
(\ref{dgkpnorm2}) are a consequence of the leptonic decay width
constraints. The parameters $\omega_{T,\,L}$ and ${\cal
N}_{T,\,L}$ are determined by solving (\ref{dgkpnorm1}) and
(\ref{dgkpnorm2}) simultaneously. In Tab. \ref{tab:1} we quote the
results for the transverse component, which it will be used in our
further analysis in the photoproduction case (longitudinal
component does not contribute at $Q^2=0$). To be consistent with
the phenomenological saturation model, which we will discuss
further, we have used the quark masses $m_{u,d,s}= 0.14$ GeV and
$m_{c}=1.5$ GeV. In the case of $\phi$ meson, we follow Ref.
\cite{dgkp:97} and take $m_s=m_{u,d}+0.15$ GeV.  We quote Ref. \cite{sandapen}
for more details in the present  approach and its comparison with data for 
 both photo and electroproduction of light mesons.

\begin{table}
\begin{center}
\begin{tabular}{||lccccc||}
\hline
\hline
$V(m_V)$ & $\hat{e}_V$ & $f_V$ & $\omega_T$  & ${\cal N}_T$ & $B_V$ \\
 MeV &  & [GeV] & [GeV] &  & [GeV$^{-2}$]  \\
\hline
\hline
$\rho\,(770)$ & $1/\sqrt{2}$ & 0.153 & 0.218 & 8.682 & 9.00 \\
$\omega\,(782)$ & $1/3\sqrt{2}$  & 0.0458 & 0.210 & 10.050 & 10.14  \\
$\phi\,(1019)$ & 1/3 & 0.079 & 0.262 & 8.000 & 8.92  \\
$J/\Psi\,(3097)$ & 2/3 & 0.270 & 0.546 & 7.665 & 4.57\\
\hline
\hline
\end{tabular}
\end{center}
\caption{Parameters and normalization
  of the  DGKP vector meson light-cone wavefunctions. Results obtained using quark mass values from the saturation model (see text). }
\label{tab:1}
\end{table}

 Finally, the imaginary part of the forward amplitude can be obtained by
 putting the expressions for photon and vector meson (DGKP) wavefunctions,
 Eqs. (\ref{wfL}-\ref{wfT}) and (\ref{dgkp_L}-\ref{dgkp_T}), into
 Eq. (\ref{sigmatot}). Moreover, summation over the quark/antiquark
 helicities  and an average over the   transverse polarization states
 of the photon should be taken into account. The longitudinal and
 transverse components are then written as \cite{sandapen, dgkp:97}
\begin{eqnarray}
{\cal I}m \, {\cal A}_{L}\!\!  & =\!\! &
 \int d^{2}\rr \int_{0}^{1} dz  \sqrt{\alpha_{\mathrm{em}}} f_{V} \,2\,z^2(1-z)^2 \,f_L(z)
\,\exp \left[\frac{-\omega_L^{2}\,\rr^{2}}{2}\right]\,Q\,K_{0}(\varepsilon r) \,\sigma_{dip}^{\mathrm{target}}(\tilde{x},\rr) \,,
\label{ampL}\\
{\cal I}m \, {\cal A}_{T}\!\! & =\!\! &  \int
d^{2}\rr \int_{0}^{1} dz \sqrt{\alpha_{{\mathrm{em}}}} \,f_{V} \,
f_T(z)\,\exp \left[\frac{-\omega_T^{2}\,\rr^{2}}{2}\right] \nonumber \\ & &
\times \,\left\{\frac{\omega_T^{2}\varepsilon r}{m_{V}}[z^{2} + (1-z)^{2}]
\,K_{1}(\varepsilon r) + \frac{m_{f}^{2}}{m_{V}}K_{0}(\varepsilon r)
\right \}\sigma_{dip}^{\mathrm{target}}(\tilde{x},\rr) \;,
\label{ampT}
\end{eqnarray}
with $\sigma_{dip}^{\mathrm{target}}$ being the dipole-proton
cross section in the nucleon case and the dipole-nucleus cross
section for scattering on nuclei. For the proton case, there are a lot of   phenomenology  for $\rho$ and $J/\Psi$ production using recent pQCD  parameterizations for the dipole-proton cross section \cite{sandapen} or considering nonperturbative QCD calculations based on stochastic QCD vacuum \cite{Dosch_Ferreira1,Dosch_Ferreira2}. In the next subsection, we briefly review the dipole-nucleon (nucleus) given by the 
phenomenological saturation model, which it will be considered in our
numerical estimates.

In order to obtain the total cross section, we assume an exponential parameterization for the small $|t|$
behavior of the amplitude. After integration over $|t|$, the total
cross section for vector meson production by real/virtual photons in
the nucleon (proton) case reads as,
\begin{eqnarray}
\sigma\, (\gamma p \rightarrow Vp) = \frac{[{\cal I}m \, {\cal A}(s,\,t=0)]^2}{16\pi\,B_V}\,(1+\beta^{2}) \;
\label{totalcs}
\end{eqnarray}
where $\beta$ is the ratio of real to imaginary part of the
amplitude and $B_V$ labels the slope parameter (we quote the
values we have used in Table \ref{tab:1}). The values considered
for the slope parameter are taken from the  parameterization used
in Ref. \cite{mara}.  For the  $\rho$ case,  we
 have taken a different value in order to describe
 simultaneously  H1 and ZEUS photoproduction data.

In addition, Eqs. (\ref{ampL}-\ref{ampT}) represent only the
leading imaginary part of the positive-signature amplitude, and its
real part can be restored using dispersion relations ${\cal R}e \,{\cal
  A}=\tan (\pi \lambda/2)\,{\cal I}m {\cal
  A}$. Thus, for the $\beta$ parameter we have used the
simple ansatz,
\begin{eqnarray}
\beta = \tan \left(\frac{\pi\lambda_{\mathrm{eff}}}{2}
\right)\,,\hspace{0.7cm} \mbox{where}\;\;\;\lambda_{\mathrm{eff}}
= \frac{\partial \,\ln\,
  [{\cal I}m \, {\cal A}(s,\,t=0)]}{\partial \,\ln s}\,,
\end{eqnarray}
with $\lambda_{\mathrm{eff}}=\lambda_{\mathrm{eff}}(W_{\gamma
N},Q^2)$  the effective power of the imaginary amplitude, which
depends on both energy and photon virtuality. The correction
coming from  real part in  photoproduction, where only
transverse component contributes, is about  3\% for light mesons
and it reaches 13\% for $J/\Psi$ at high energies. It is worth mentioning that a different computation of the $\beta$ parameter, as in Ref. \cite{sandapen}, produces a larger effect even in the photoproduction case. An
additional correction is still required for heavy mesons, like
$J/\Psi$. Namely,  skewedness effects which takes into account the
off-forward features of the process (different transverse momenta
of the exchanged gluons in the $t$-channel),   are increasingly
important in this case.  Here, we follow  the studies in Ref.
\cite{Shuvaev:1999ce}, where the ratio of off-forward to forward
gluon  distributions reads as \cite{Shuvaev:1999ce},
\begin{eqnarray}
R_{g}\,(\lambda_{\mathrm{eff}})=\frac{2^{2\lambda_{\mathrm{eff}} + 3}}{\sqrt{\pi}}\,\frac{\Gamma\,\left(\lambda_{\mathrm{eff}}+ \frac{5}{2}\right)}{\Gamma \,\left(\lambda_{\mathrm{eff}}+4 \right)}\,,
\label{skew}
\end{eqnarray}
and we will multiply the total cross section by the factor $R_g^2$ for
the heavy meson case.

In the case of nuclear targets, $B_V$ is dominated by the nuclear
size, with $B\sim R_A^2$ ($R_A=1.2\,A^{1/3}$ fm is the nuclear
radius) and the non-forward differential cross section is
dominated by the nuclear form factor, which is the Fourier
transform of the nuclear density profile. Here we use the
analytical approximation of the Woods-Saxon distribution as a hard
sphere, with radius $R_A$, convoluted with a Yukawa potential with
range $a=0.7$ fm. Thus, the nuclear form factor reads
as \cite{klein},
\begin{equation}
F(q=\sqrt{|t|}) = \frac{4\pi\rho_0}{A\,q^3}\,
\left[\sin(qR_A)-qR_a\cos(qR_A)\right]
\,\left[\frac{1}{1+a^2q^2}\right]\,\,, \label{FFN}
\end{equation}
where $\rho_0 = 0.16$ fm$^{-3}$.

 The photonuclear cross section is given by
\begin{eqnarray}
\sigma\,(\gamma A\rightarrow VA) =  \frac{[{\cal I}m \, {\cal
      A}_{\mathrm{nuc}}(s,\,t=0)]^2}{16\pi}\,\,(1+\beta^{2})\,\int_{t_{min}}^\infty
      dt\, |F(t)|^2 \,,
\label{fotonuclear}
\end{eqnarray}
with $t_{min}=(m_V^2/2\,\omega)^2$, where $\omega$ is the photon energy.

 Having introduced the main expressions
for  computing vector meson production in the color dipole
approach, in what follows we present the saturation model and its
extension for the scattering on nuclei targets.

\subsection{Dipole-nucleus cross section in the saturation model}

For electron-proton interactions, the dipole  cross section
$\sigma_{dip}^{\mathrm{proton}}$, describing the dipole-proton
interaction, is substantially affected by nonperturbative content.
There are several phenomenological implementations for this
quantity. The main feature of these approaches is
to be able to match the soft (low $Q^2$) and hard (large $Q^2$)
regimes in an unified way. In the present work, we follow the
quite successful saturation model \cite{golecwus}, which
interpolates between the small and large dipole configurations,
providing color transparency behavior, $\sigma_{dip}\sim \rr^2$,
as $\rr \rightarrow 0$ and constant behavior, $\sigma_{dip}\sim
\sigma_0$, at large dipole separations. The parameters of the
model have been obtained from an adjustment to small $x$ HERA
data. Its parameter-free application to diffractive DIS has been
also quite successful \cite{golecwus} as well as its extension to
virtual Compton scattering \cite{Favart_Machado}, vector meson
production \cite{mara, sandapen} and two-photon collisions
\cite{Kwien_Motyka}.   The parameterization for the dipole cross
section takes the eikonal-like form \cite{golecwus},
\begin{eqnarray}
\sigma_{dip}^{\mathrm{proton}} (\tilde{x}, \,\rr^2)  =  \sigma_0 \,
\left[\, 1- \exp \left(-\frac{\,Q_{\mathrm{sat}}^2(\tilde{x})\,\rr^2}{4} \right) \, \right]\,, \hspace{1cm} Q_{\mathrm{sat}}^2(\tilde{x})  =  \left( \frac{x_0}{\tilde{x}}
\right)^{\lambda} \,\,\mathrm{GeV}^2\,,
\label{gbwdip}
\end{eqnarray}
where the saturation scale $Q_{\mathrm{sat}}^2$ defines the onset of the
saturation phenomenon, which depends on energy. The parameters, obtained from a fit to the small-$x$ HERA data, are $\sigma_0=23.03 \,(29.12)$ mb, $\lambda= 0.288 \, (0.277)$ and
$x_0=3.04 \cdot 10^{-4} \, (0.41 \cdot 10^{-4})$ for a 3-flavor
(4-flavor) analysis. An
additional parameter is the effective light quark mass, $m_f=0.14$
GeV, which plays the role of a regulator for the photoproduction
($Q^2=0$) cross section, as discussed before.  The charm quark mass is
considered to be $m_c=1.5$ GeV. A smooth transition to the
photoproduction limit is obtained via the scaling variable \cite{golecwus},
\begin{eqnarray}
\tilde{x}= \frac{Q^2 + 4\,m_f^2}{Q^2 + W_{\gamma N}^2} \,.
\end{eqnarray}
The
saturation model is suitable in the region below $x=0.01$ and the
large $x$ limit needs still a consistent  treatment. Making use of
the dimensional-cutting rules, here we supplement
 the dipole cross section, Eq. (\ref{gbwdip}), with a threshold factor
$(1-x)^{n_{\mathrm{thres}}}$, taking $n_{\mathrm{thres}}=5$ for a
3-flavor analysis and $n_{\mathrm{thres}}=7$ for a 4-flavor one.
This procedure ensures consistent description of heavy quark
production at the fixed target data \cite{Mariotto_Machado}.

Let us discuss the extension of the saturation  model for the
photon-nucleus interactions. Here, we follow the simple procedure
proposed in Ref. \cite{armesto}, which consists of an extension to
nuclei of the
saturation model discussed above, using the Glauber-Gribov picture \cite{gribov}, without any new parameter. In this
approach, the nuclear version is obtained replacing the
dipole-nucleon cross section in Eq. (\ref{sigmatot}) by the
nuclear one,
\begin{eqnarray}
\sigma_{dip}^{\mathrm{nucleus}} (\tilde{x}, \,\rr^2;\, A)  = 2\,\int d^2b \,
\left\{\, 1- \exp \left[-\frac{1}{2}\,T_A(b)\,\sigma_{dip}^{\mathrm{proton}} (\tilde{x}, \,\rr^2)  \right] \, \right\}\,,
\label{sigmanuc}
\end{eqnarray}
where $b$ is the impact parameter of the center of the dipole
relative to the center of the nucleus and the integrand gives the
total dipole-nucleus cross section for a  fixed impact parameter.
The nuclear profile function is labeled by $T_A(b)$, which will
be obtained from a 3-parameter Fermi distribution for the nuclear
density  \cite{devries}. The above equation sums up all the
multiple elastic rescattering diagrams of the $q \overline{q}$
pair and is justified for large coherence length, where the
transverse separation $r$ of partons in the multiparton Fock state
of the photon becomes as good a conserved quantity as the angular
momentum, {\it i. e.} the size of the pair $r$ becomes eigenvalue
of the scattering matrix. It is important to emphasize that for
very small values of $x$, other diagrams beyond the multiple
Pomeron exchange considered here should contribute ({\it e.g.}
Pomeron loops) and a more general approach for the high density
(saturation) regime must be considered. However, we believe that 
this approach allows us to obtain lower limits of the high density 
effects at eRHIC and HERA-A. Therefore, at first glance, the region of applicability of this  model should be at small values of  $x$, i.e. large coherence length, and for
not too high  values of virtualities, where the implementation of
the DGLAP evolution should be required. Consequently, the approach
is quite suitable for the analysis of exclusive vector meson
photoproduction in the   kinematical range of the planned
lepton-nucleus colliders (eRHIC and HERA-A). Furthermore, it should be noticed that the energy dependence of the cross sections is strongly connected with the  semi-hard
scale (the saturation momentum scale). Namely, the saturation effects are larger  whether the momentum scale is of order or larger than the correspondent size
 of the vector meson and the energy growth of the cross section is then slowed down.

\section{Vector meson photoproduction in the dipole Pomeron framework}
\label{sec:dipolePom}

Let us summarize the main features and expressions for the
nonperturbative approach given by the dipole Pomeron model \cite{dipolepom}. This
model describes the vector meson exclusive photoproduction \cite{martynov_predazzi} data
from HERA without need of a Pomeron contribution with intercept
higher than 1, thus not violating the Froissart-Martin bound.  The
picture of the interaction is given by a photon fluctuating into a
quark-antiquark pair  and further the nucleon (proton) interacts
with it through Pomeron or secondary Reggeon exchange. After that,
the pair converts into a vector meson. In general lines, this
picture is quite similar to that one for interaction among hadrons
in the Regge limit. In particular for photoproduction, the
representation of the  photon as a hadron is reasonably supported
and the  Regge pole theory, with a Pomeron
 universal in all hadron-hadron interactions, can be safely used
 there.

For the  Pomeron contribution we follow Ref. \cite{martynov_predazzi} and one considers
the dipole Pomeron, which gives a very good description of all
hadron-hadron total cross sections. As Pomeron and secondary Reggeons would be universal objects
 in Regge theory, the corresponding $j$-singularities of photon-proton
 amplitudes and their trajectories at the photoproduction limit
 coincide with those appearing in pure hadronic amplitudes. In
 particular, the restriction on the Pomeron intercept implied by the
 Froissart-Martin bound suggests that it is a more complicated
 singularity instead of a simple pole having a universal intercept
 $\alpha_{\pom}\geq 1$, which one would apply also to DIS. In the case
 of a dipole Pomeron, it is a double $j$-pole leading to
 $\sigma_{tot}^{hh,\,\gamma h}\propto \ln s$ and unitarity requirements are covered.

Let's consider the usual Mandelstam variables, $s=W_{\gamma
  N}^2=m_N^2+2m_N\nu-Q^2$ ($m_N$ is the nucleon mass), $t$ (momentum transfer) and
defining the scaling variable, $\overline{Q}^2=Q^2+m_V^2$. Making use
  of the  latter quantity, the dipole Pomeron model can be generalized for
  virtual external particles. The scattering amplitude is given by the
  contribution of Reggeons at low energies and the dipole Pomeron
  ($\alpha_{\pom}(t=0)=1$) dominates at higher energies. A simple pole
  parameterization is used for the $f$-Reggeon. The parameters
  $\alpha_{\pom}(t)$ and $\alpha_{I\!\!R}(t)$ are universal and do not depend of the
  reaction, whereas couplings $g_i$, energy scales $s_{0\,i}$ and
  slopes $b_i$ are functions
  of the scaling variable and the same for all reactions.

Taking into account the features discussed above, the differential elastic cross section is written as \cite{martynov_predazzi}
\begin{eqnarray}
\frac{d\sigma}{d\,t} \,(\gamma\, p\rightarrow
Vp) = 4\pi\, \left|\,{\cal
    A}_{\pom}\,(s,t\,;\,m_V^{2}) + {\cal
    A}_{I\!\!R}\,(s,t\,;\,m_V^{2}) \,\right|^{2}\;,
\label{sigmaregge}
\end{eqnarray}
where in the photoproduction case the amplitudes for the  secondary
Reggeons and Pomeron contributions are parameterized as \cite{martynov_predazzi},
\begin{eqnarray}
\label{pomeron}
{\cal A}_{\pom}(W_{\gamma N}^2,t\,;\,m_V^2) & = & i\,g_0(t;
m_V^2)\,\left(-i\,\frac{W_{\gamma N}^2-m_p^2}{W_0^2+m_V^2}
\right)^{\alpha_{\pom} (t)-1} \nonumber \\
& & +\,\, i\,g_1(t; m_V^2)\,\ln
\left(-i\,\frac{W_{\gamma N}^2-m_p^2}{W_0^2+m_V^2}\right)
\left(-i\,\frac{W_{\gamma
    N}^2-m_p^2}{W_0^2+m_V^2}\right)^{\alpha_{\pom} (t)-1}\\
{\cal A}_{I\!\!R}\, (W_{\gamma N}^2,t;\,m_V^2) & = &
i\,g_{I\!\!R}\,(t;\,m_V^2)\,\left(-i\,\frac{W_{\gamma N}^2-m_p^2}
{W_0^2+m_V^2}
\right)^{\alpha_{I\!\!R}(t)-1}\,,
\end{eqnarray}
 where one takes a linear Pomeron trajectory $\alpha_{\pom}
 (t)=1+\alpha^{\prime}_{\pom} (0)\, t$,  with the usual value for the
 slope $\alpha^{\prime}_{\pom} (0)=0.25$ GeV$^{-2}$. The Reggeons and
 Pomeron couplings are written as,
\begin{eqnarray}
g_{\,0,1}\,(t\,; \,m_V^2) & = & \frac{g_{\,0,1} \,m_V^2}{(W_0^2+m_V^2)^2}\,\exp
\left(b_{\pom}^2 \,t \right)\,, \\
g_{I\!\!R}\,(t\,;\, m_V^2) & = & \frac{g_{I\!\!R}\,
  m_p^2}{(W_0^2+m_V^2)^2}\,\exp \left (b_{I\!\!R}^2\, t\right)\,,
\end{eqnarray}
where the couplings $g_{\,0,1}$,  the energy scale $W_0^2$
(GeV$^2$) and $t$-slope $b^2_{\pom}$ (GeV$^{-2}$) for the Pomeron are
adjustable parameters of the model; in addition $m_p^2$ is the
proton mass. The notation $V$ stands for $\rho$, $\phi$ and
$J/\psi$, whereas $I\!\!R=f,\pi$ for $\omega$. The remaining
constants for the Reggeons, $g_f$, $g_\pi$, $b^2_{I\!\!R}$
(GeV$^{-2}$) are also adjustable parameters. One uses the same
slope $b^2_{I\!\!R}$ for $f$ and $\pi$ Reggeon exchanges. We quote
Ref. \cite{martynov_predazzi} for details on the fit procedure and the tables for the 
fitted parameters. Let us stress that the only variable that
differentiates among the various vector meson elastic cross
sections is the mass of the vector mesons.

Following Ref. \cite{martynov_predazzi}, some additional comments are in order. The
behavior in the threshold region is given by multiplying the
amplitudes by the correction factor $(1-\tilde{x})^{\delta}$, where
$\tilde{x}= (m_p+m_V)^2/W_{\gamma N}^2$ and  $(m_p+m_V)$ is the reaction
threshold. The power $\delta=\sqrt{m_V^2/m_0^2}$ drives the energy
dependence in that region, with $m_0^2$ (GeV$^2$) fitted from data.
Concerning details in Ref. \cite{martynov_predazzi} when taking into account the secondary Reggeons, for $\rho$, $\phi$ and $J/\psi$ meson photoproduction  the
scattering amplitude was written as the sum of a Pomeron and $f$ contribution.
Although according  to Okubo-Zweig rule, the $f$ meson contribution ought to be
suppressed in the production of $\phi$ and $J/\psi$ mesons, the  $f$
meson contribution was added even in the $J/\psi$ meson case. For
$J/\psi$, it was found to be negligible whereas it is sizable  for $\phi$ meson production.

The $\gamma p \rightarrow Vp$ process can be used as input in the
calculations of the total cross section for the reaction $\gamma A
\rightarrow V A$. A major simplification comes from the use of
vector meson dominance, which allows to relate this
photoproduction cross section to the cross section for the forward
elastic $Vp \rightarrow Vp$ scattering. Following vector meson
dominance \cite{review,gilman},
\begin{eqnarray}
{d\sigma(\gamma p\rightarrow Vp) \over dt} \bigg|_{t=0} =
{4\pi\alpha_{\mathrm{em}} \over f_V^2} \ \ {d\,\sigma (Vp\rightarrow Vp)\over dt}
\bigg |_{t=0}\,\,.
\end{eqnarray}
where, t is the squared 4-momentum transfer between the proton and
vector meson, $\alpha_{\mathrm{em}}$ is the electromagnetic coupling constant and $f_V$ is the vector meson-photon coupling, $f_V = 4\,\pi\,m_V\alpha_{\mathrm{em}}^2/ (3\,\Gamma_{V\rightarrow e^+e^-})$, with $m_V$ the vector meson mass and $\Gamma_{V \rightarrow
e^+e^-}$ the leptonic decay partial width.   Values for
$f_V^2/4\pi$ are given in Table II of Ref. \cite{klein}. Using the
Optical theorem, the total cross section is given by, 
\begin{eqnarray}
\sigma_{tot}(Vp \rightarrow Vp) = \left[16\,\pi\, \frac{d\,\sigma(Vp\rightarrow Vp)}{dt}
\bigg|_{t=0}\right]^{\frac{1}{2}}.
\end{eqnarray}
 The scattering cross
section from heavy nuclei can be found by a (quantum mechanical)
Glauber-Gribov calculation,
\begin{equation} \sigma_{tot}(VA \rightarrow VA) =
2\int d^2 b \,\left[1 - \exp \left(-\frac{1}{2}\,T_{A}(b)\,\sigma_{tot}(Vp\rightarrow Vp)\right)\right].
\end{equation}
As referred before, the nuclear profile function is labeled by $T_A(b)$, which will
be obtained from a 3-parameter Fermi distribution for the nuclear
density  \cite{devries}. The Optical theorem for nucleus $A$ and
vector meson dominance are then used to find the following relation,
\begin{eqnarray}
\frac{d\,\sigma(\gamma A\rightarrow VA)}{dt} \bigg|_{t=0} =
\frac{\alpha_{\mathrm{em}}}{4f_V^2} \,\sigma_{tot}^2(VA \rightarrow VA)\,\,.
\end{eqnarray}
From this equation one can directly understand the $A$-dependence in two
limiting cases: in the transparent limit there is a $A^2$ behavior
(typical of coherent processes) and in the black disc limit we
have an $A^{\frac{4}{3}}$ rise with the nuclear number $A$.
 The total
photonuclear cross section is then given by
\begin{eqnarray}
\sigma_{tot}(\gamma A\rightarrow VA) = {d\sigma(\gamma A\rightarrow
VA)\over dt}\bigg|_{t=0} \int_{t_{min}}^\infty dt |F(t)|^2\,\,,
\label{Reggenuc}
\end{eqnarray}
where $F(t)$ is given in Eq. (\ref{FFN}). As $F(t)$ is
$A$-dependent we have that integration over $t$ yields a factor of
$A^{-\frac{2}{3}}$, which implies a $A^{\frac{4}{3}}$
($A^{\frac{2}{3}}$) behavior in the transparent (black disc)
limit.

\section{Nuclear vector meson exclusive photoproduction}
\label{photo}

 In this section we compute the nuclear cross section for the
 exclusive photoproduction of vector mesons. Here, we compare the QCD
 approach given by the saturation model extended to nuclei targets
 within the  color dipole picture  as well as a nonperturbative approach rendered by the dipole Pomeron model,
 which does also not violate unitarity at high energies. We focus on the
 energy range and nuclei targets expected for the future lepton-nuclei
 colliders (eRHIC and eHERA) and also for available range to be covered in
 ultraperipheral heavy-ion collisions (UPC's) at LHC.

\begin{figure}[t]
\begin{tabular}{cc}
\psfig{file=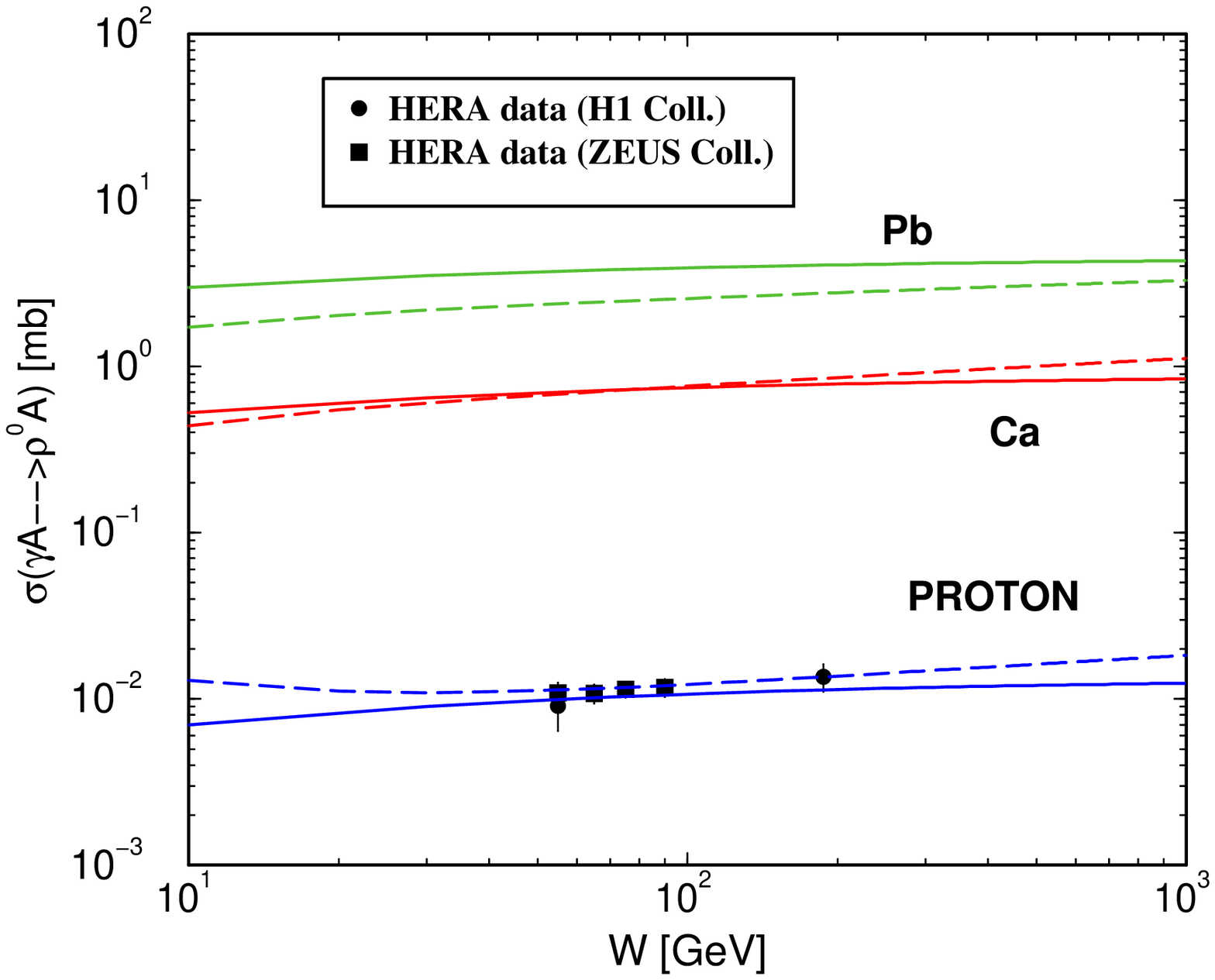,width=80mm} &
\psfig{file=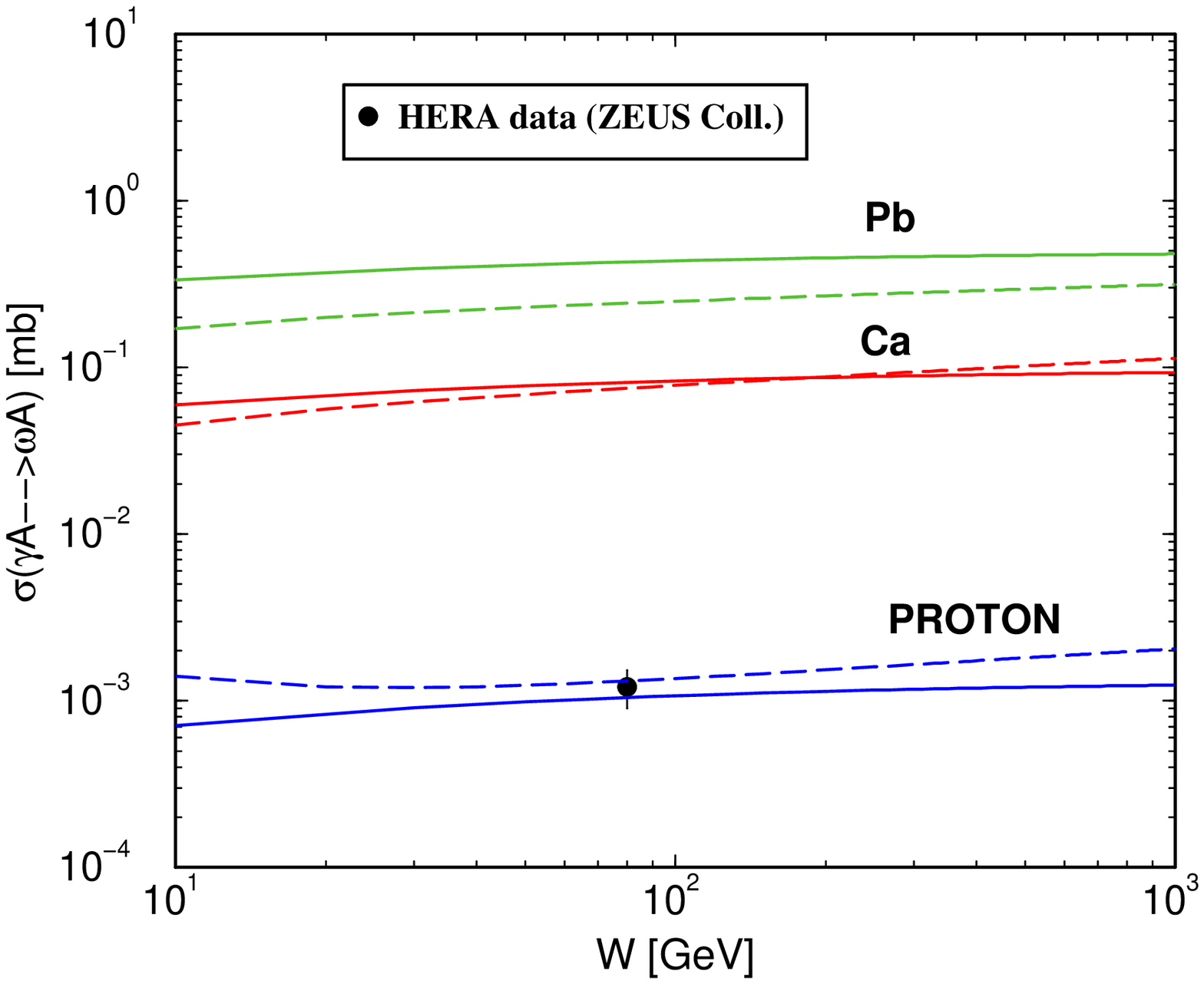,width=77mm}
\end{tabular}
 \caption{\it The total cross section for $\rho$ and $\omega$
 photoproduction on proton as well as for light (Ca) and heavy (Pb)
 nuclei. The solid lines stand for the QCD color dipole approach
 and the dashed  ones for the soft dipole Pomeron approach. Experimental
  high energy data from DESY-HERA collider on proton target are also shown.}
\label{fig1}
\end{figure}

\begin{figure}[t]
\begin{tabular}{cc}
\psfig{file=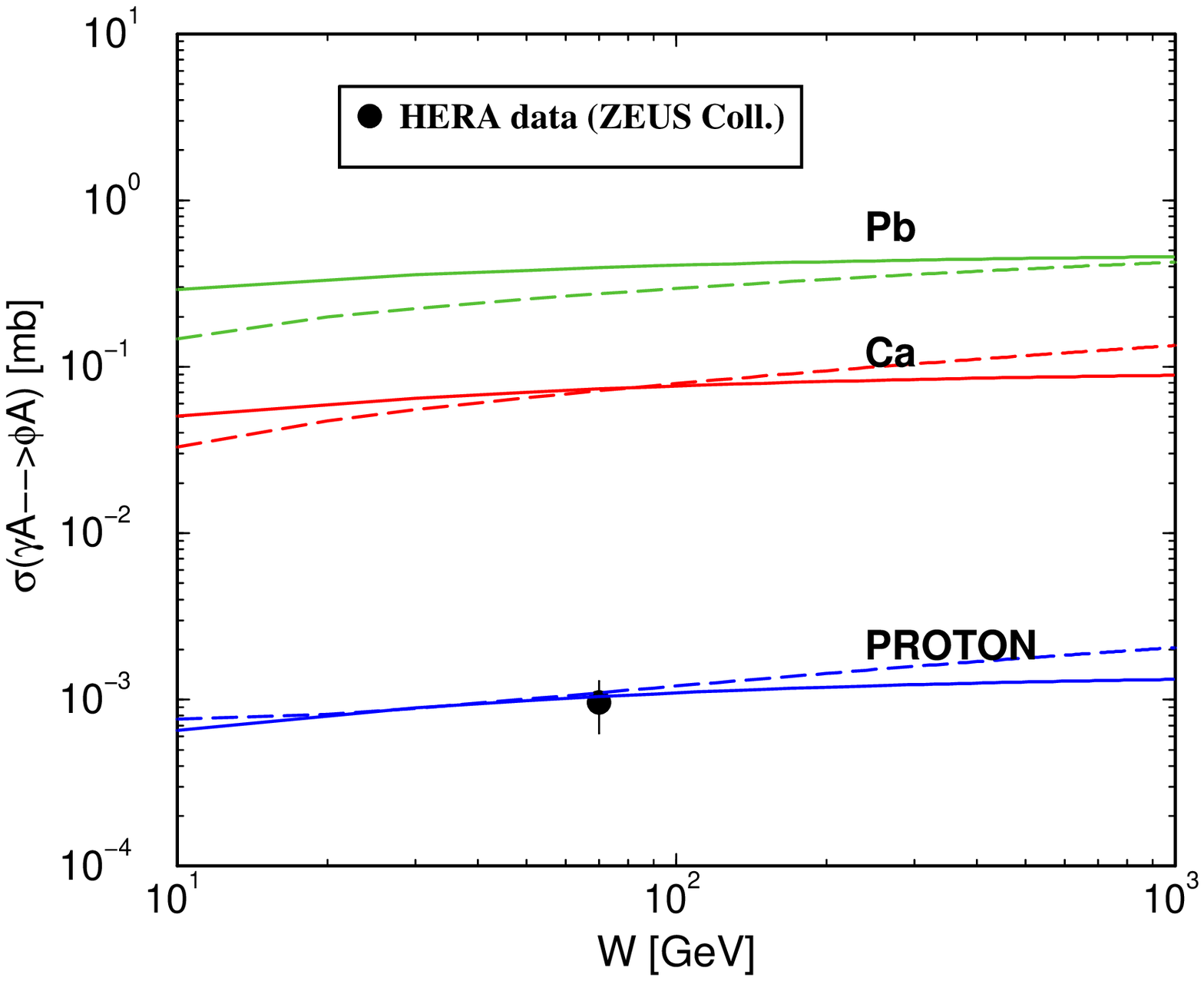,width=80mm} &
\psfig{file=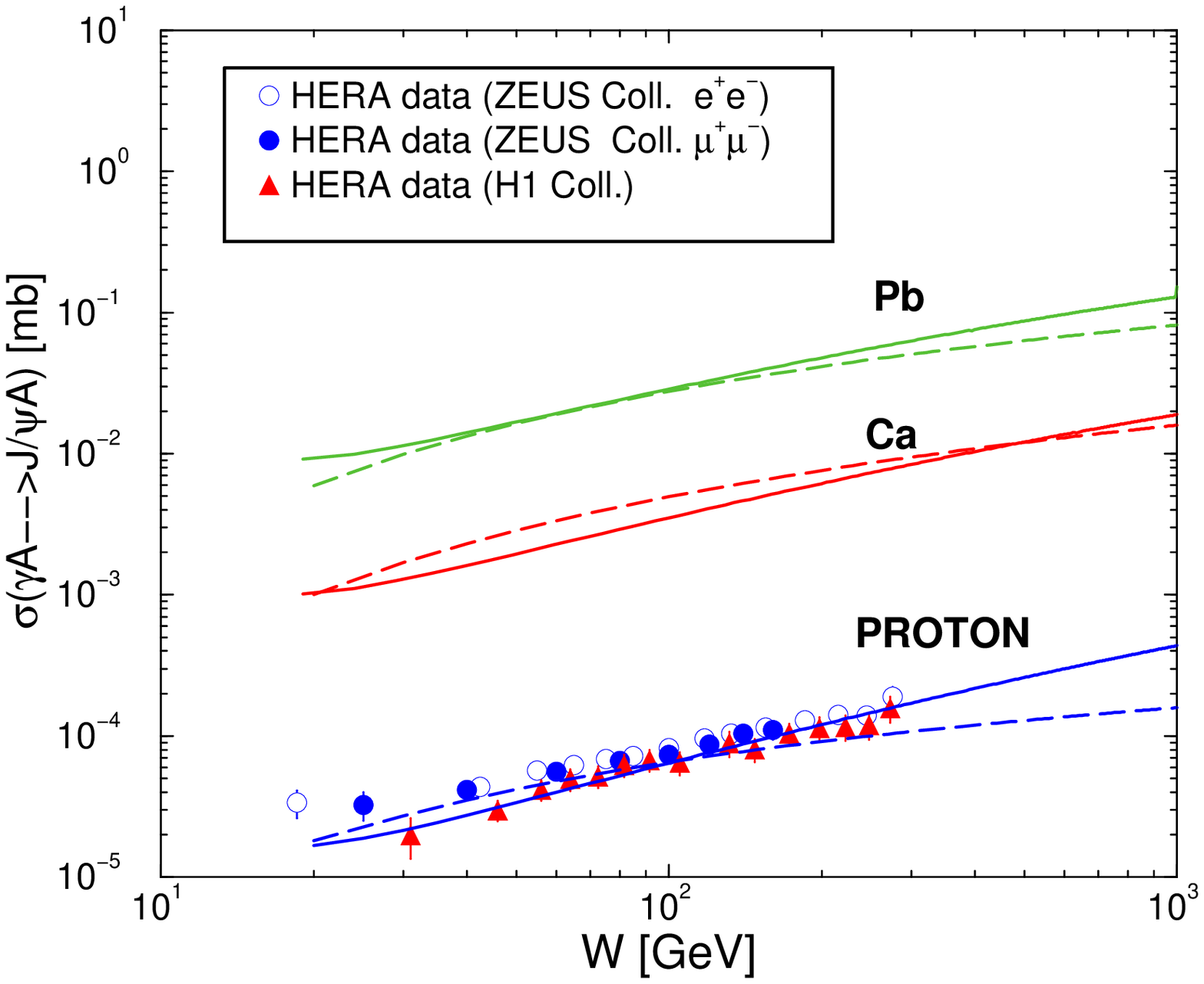,width=77mm}
\end{tabular}
 \caption{\it The total cross section for $\phi$ and $J/\Psi$
 photoproduction on proton as well as for light (Ca) and heavy (Pb)
 nuclei. The  solid lines stand for the QCD color dipole approach
 and the dashed  ones for the soft dipole Pomeron approach. Experimental
  high energy data from DESY-HERA collider on proton target are also shown. }
\label{fig2}
\end{figure}

 In Figs. \ref{fig1} and \ref{fig2}
are shown the results for the $\rho$,  $\omega$, $\phi$ and
$J/\Psi$ photoproduction cross section as a function of energy for
different nuclei, including the proton case.
 The results in the pQCD
and Regge approaches present mild growth on $W_{\gamma N}$ at high
energies stemming from the high energy behavior of the models,
whereas the low energy region is consistently described through
the threshold factor.  For the proton, the experimental data
from DESY-HERA collider \cite{ZEUS_rho,H1_rho,ZEUS_phi,ZEUS_omega,H1_jpsi,ZEUS_jpsi}  are also included for sake of
comparison. We can see that the $\rho$, $\omega$ and $\phi$
results vary only slowly with energy, in contrast with the
$J/\Psi$ predictions. The saturation model (solid lines) gives a flatter energy dependence in comparison with the dipole Pomeron model (dashed lines). On the other hand, for $J/\Psi$ the situation changes, where the dipole Pomeron model producing a mild increasing at high energies in comparison with the saturation model.

Let us discuss the results coming from the color dipole approach. For the proton
case we have used Eqs. (\ref{ampT}-\ref{totalcs}) and the
dipole-proton cross section given by Eq. (\ref{gbwdip}). The
contribution of the real part of amplitude is small for the light
mesons, whereas is sizeable for the $J/\Psi$ case. Moreover, the
 skewedness correction to the $J/\Psi$ photoproduction is important,
 providing a larger overall normalization, as discussed in
 Sec. \ref{sec:dipolePic}. These features remain in the nucleus case,
 where we have used Eq. (\ref{fotonuclear}) and the dipole-nucleus
 cross section given by Eq. (\ref{sigmanuc}). It is worth mentioning
 that the effective power of the imaginary part of amplitude is slowed
 down in the nuclear case and this has implications in the corrections
 of real part and skewedness. The results for photonuclear production
 on nuclei is consistent with the studies in Ref. \cite{klein}, except for
 $J/\Psi$ once the growth on energy is mild at high energies in the
 present  case. The agreement with the proton data is  consistent
 and the extension to Ca and Pb targets is suitable since it is
 constrained by the DESY-HERA data and validity of the model in the
 energy range considered. Furthermore, the present investigation is
 complementary to those ones on heavy quark production \cite{vicmagsat} and
 nuclear structure functions \cite{armesto} using the saturation model for 
 nuclear targets.

 For light mesons production in $\gamma p$ processes, the dipole
Pomeron model predicts a larger growth with the energy than the
saturation model due to the dominance
of large $q\overline{q}$ pair separations in the saturation
approach. This implies that $\sigma_{dip} \propto \sigma_0$, i.e.
almost energy independent. It can be checked that the integration over dipole sizes in this dipole configuration gives an almost constant value, without logarithmic corrections as in DIS case.   Differently, the dipole Pomeron model
predicts a logarithmic dependence in the energy. In the nuclear
case, this behavior implies a larger modification of the cross
section in the dipole Pomeron model in comparison with the
saturation model. In particular, we have that for the nuclear
exclusive vector meson photoproduction with $A$ = Pb we predict
that the difference between the results of the models should be a
factor about 1.5. In contrast, for $J/\Psi$ photoproduction, we have that
the saturation model predicts a larger cross section for high
energies. This is associated to the  color transparency regime,
present due to the small pair separation between charm and
anti-charm. In this case we have a power-like behavior in contrast
with a logarithmic dependence present in the dipole Pomeron model.

A final comment on the small-$t$ approximation considered here is in order. As shown in Ref. \cite{Kowalski:2003hm}, the saturation effects play an important role in the $t$ dependence of the scattering amplitude, mostly at large $t$. Therefore, the approximation presented here should be justified. Our master equation is exactly the same as in Refs. \cite{Kowalski:2003hm,stasto}, which reads,
\begin{equation}
  \frac{d\sigma_{L,T}}{dt} = \frac{1}{16\pi} \left|
  \int d^2 {\bf r} \int \frac{dz}{4\,\pi} \int d^2b\, 
  \left(\Psi_V\,\Psi_{\gamma}\right)_{L,T}\, e^{-i {\bf b}\cdot{\bf \Delta}}\, 
  \frac{d\sigma_{qq}}{d^2b}\, \right|^2
\label{teaney}
\end{equation}
where the squared momentum transfer is denoted by $\Delta^2 = -t$.  Basically, our expressions Eqs. (\ref{fotonuclear}) and (\ref{Reggenuc}) stand for the differential cross section at $t=0$  and further we have used the standard approximations for the small $t$ behavior of the scattering amplitude on nucleon (proton) and nuclei. Namely, for the proton case one has considered the usual exponential parameterization which includes  the meson slope parameter $B_V$.  For the scattering on nuclei, this is accounted for by the nuclear form factor $F_1(t)$, which includes the correct size of the nuclear target. Considering the saturation model,  $q\bar{q}$ differential cross section is given by,
\begin{equation}
\frac{d\sigma_{q\bar{q}}\,(x,\rr,b)}{d^2b} = 2\,\left[1-
\exp\left(-\frac{1}{2}\,\sigma_{dip}(x,\rr)\,T(b)\right)
\right]  \; .                 
\end{equation}

The remaining issue is what the accuracy of such an  approximation concerning the saturation region for DIS with a nucleon/nucleus target. This can be addressed by looking at the Fourier transform of the differential dipole cross section for large dipole sizes (saturation limit). It was shown in Ref.  \cite{Kowalski:2003hm} [see Figs. (16) and (17) in that paper], the saturation effects predict diffractive dips at large $t$. However, the pQCD picture remains quite the same at small $t$. Therefore, we believe that the integration on $t$ of the complete expression, Eq. (\ref{teaney}), should be not strongly sensitive to the large $t$ region, once it is sub-dominant in the whole integrand. That is, we expect the effect on the total cross section is hidden in the integration, whereas it is important at large $t$ in the differential cross section.

\section{Summary and Conclusions}
\label{conc}

\begin{figure}[t]
\begin{tabular}{cc}
\psfig{file=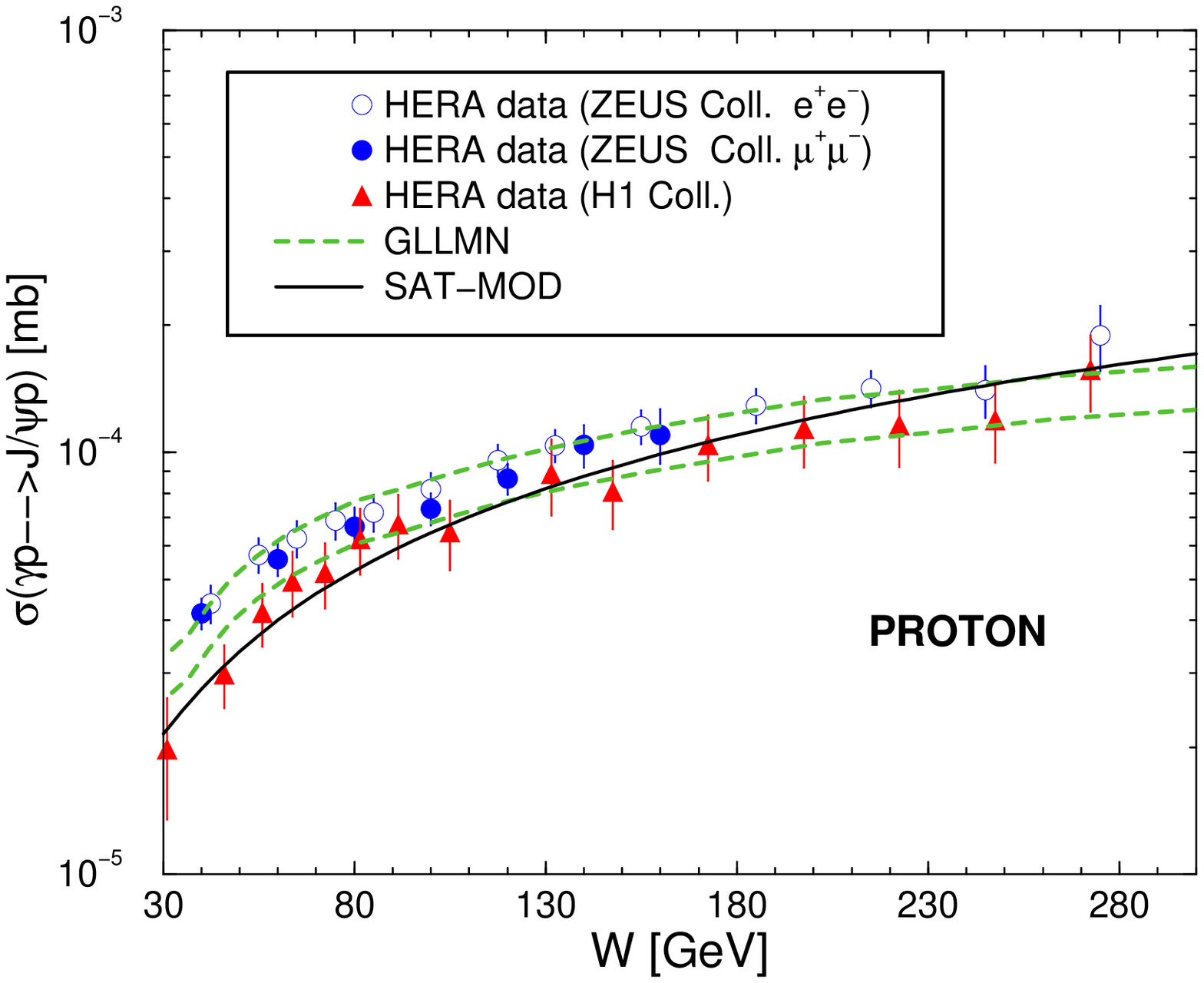,width=80mm} &
\psfig{file=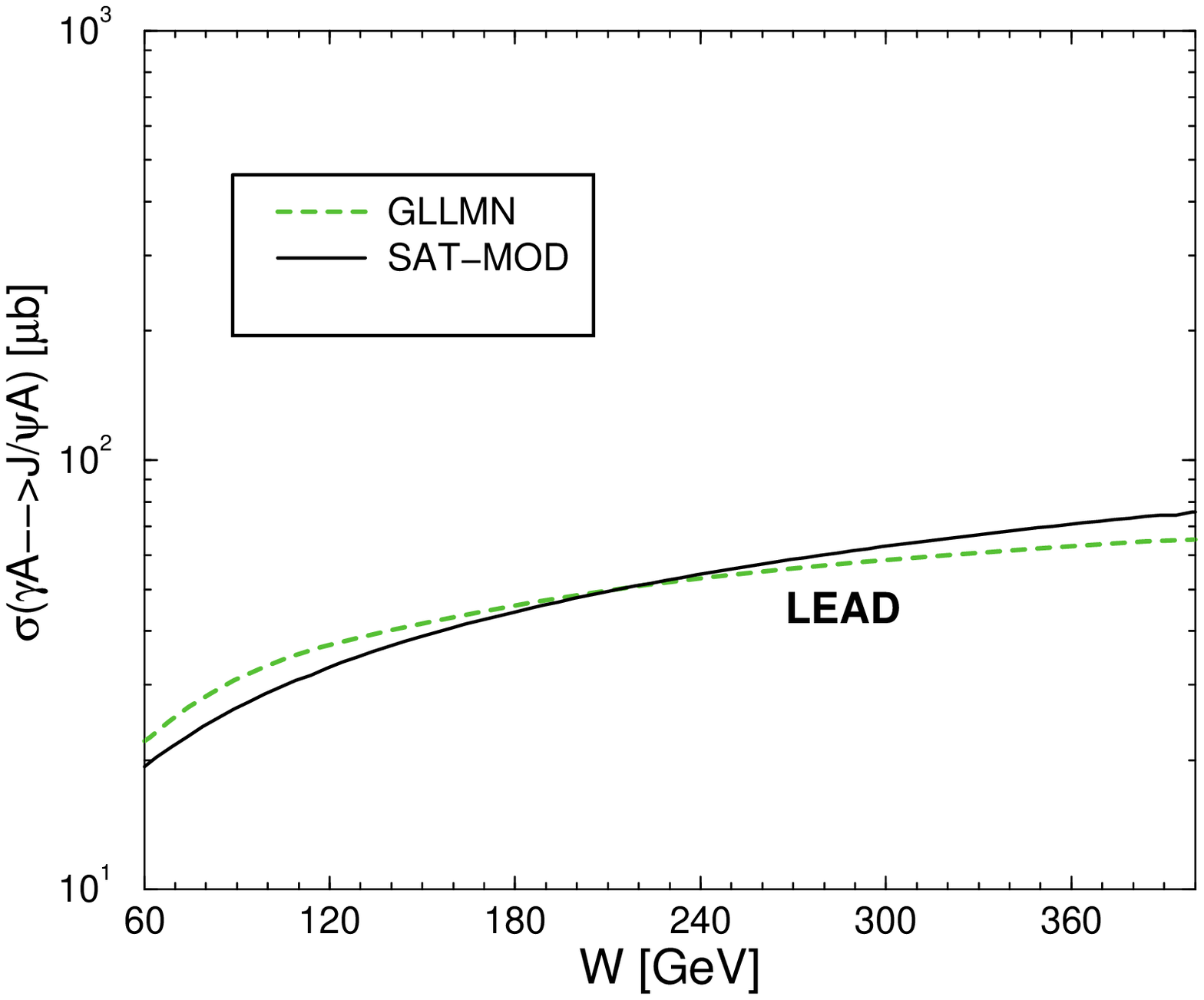,width=77mm}
\end{tabular}
 \caption{\it The total cross section for $J/\Psi$
 photoproduction for  proton target (left panel) and for  lead nucleus (right panel). The  solid lines stand for the saturation model (SAT-MOD)  and the dashed  ones for the GLLMN model \cite{Gotsman:2003ww}. }
\label{fig3}
\end{figure}

In this paper we have calculated the nuclear  cross sections for
exclusive vector meson  photoproduction within  QCD  color
dipole  picture and Regge approach. These models predict cross
section which does not violate the Froissart-Martin bound at high
energies. Since they describe reasonably the experimental data for
nucleon (proton) target, we are confident in extending these
models for the nuclear photoproduction  case.

For the first approach, we have considered the saturation
 model, which is analytically simple and  gives a good description of inclusive and
diffractive $ep$ experimental data. This model should be valid
until the full non-linear  evolution effects become  important,
which implies the consideration of the Pomeron loops beyond the
multiple scattering on single nucleons estimated in the present
framework. We have verified that the energy behavior is mild, mostly
 for $J/\Psi$ where we would expect a hard behavior. This means that
 an important contribution also comes from large dipole
 configurations, related to soft domain.   We predict absolute values
 for the cross section rather large, being about 4 mb and 0.12  mb for
 $\rho$ and $J/\Psi$, respectively, for lead at  $W_{\gamma N} \approx 1$ TeV. These values are
similar to those resulting from Ref. \cite{klein}, except for the mild
 energy behavior for $J/\Psi$
 case presented here.  Concerning the
$A$-dependence, we have found a behavior proportional to $A^{2/3}\,(A^{4/3})$
for the  cross sections of light (heavy) mesons,  in agreement with
theoretical expectations associated with a transition to  the
black disc regime. In fact, for light mesons the nuclear shadowing in the scattering amplitude is stronger than a simple $A^{1/3}$ ansatz, as discussed in Ref. \cite{armesto} when computing the nuclear structure functions. Namely, the saturation scale, which drives the $A$-dependence, between the proton and central nucleus is not simply $\propto A^{1/3}$, but has a prefactor which makes the result smaller.

The results presented here can be contrasted, at least for the $J/\Psi$ case, with the results of Ref. \cite{Gotsman:2003ww} (hereafter GLLMN model), where the vector meson production (including DIS production) has been addressed in the color dipole picture. That analysis considers the  Glauber approach and a numerical solution of the Balitsky-Kovchegov nonlinear evolution equation for the imaginary part of the dipole-nucleon scattering amplitude \cite{Gotsman:2002yy}. The comparison is shown in Fig. (\ref{fig3}), where the solid lines represent the results from the saturation model (SAT-MOD) and the dashed lines are the numerical result of the GLLMN model. For the proton case (left panel), the saturation model gives a steeper growth on energy, whereas GLLMN produces a mild behavior at large energies. The upper/lower GLLMN curves stand for maximum and minimum values for the total cross section, obtained considering two different values for the correction factor $K_F$ (we quote Ref. \cite{Gotsman:2003ww} for more details). It should be noticed that the  behavior near threshold is also  different in the two models. For the nucleus case (right panel), the behavior on energy remains basically the same as for the proton. The GLLMN model gives somewhat a slightly lower  cross section at large energies, which it is about 8\% smaller than the result for the saturation model at $W_{\gamma A}\approx 400$ GeV and presents a flatter behavior on energy.  These features are directly associated with the different dipole-nucleon cross sections used in the two approaches. However, considering the relative errors of order of 15-20\% in the predictions from Ref. \cite{Gotsman:2003ww}, as stated by the authors, we can conclude that the agreement between the results is satisfactory, despite the distinct approximations made in the calculations.

Concerning the nonperturbative approach, we have considered an unitarized Pomeron model and computed consistently its extension for nuclear targets. The energy dependence is logarithmic in any case and the nuclear effects seem to be stronger than in the saturation model for  heavy nuclei. 
The nuclear dependence follows similar behavior as for the saturation model. However, it should be noticed we have used a quantum mechanical Glauber-Gribov calculation for nuclei targets in contrast with the results presented in Ref. \cite{klein}. As pointed out in Ref. \cite{Frankfurt:2002sv}, the present procedure gives a cross section higher than the classical mechanical model used for the predictions in Ref. \cite{klein}. For instance, it has been found in \cite{Frankfurt:2002sv} a difference by  a factor  2.5 in $\rho$ photoproduction at RHIC energies.

Our results demonstrate that the experimental analyzes of  nuclear
exclusive vector meson photoproduction in the future
electron-nucleus colliders eRHIC and HERA-A could be useful to
discriminate between the different theoretical scenarios, mainly
if heavy nuclei are considered. An alternative until these
colliders become reality is the possibility of using
ultraperipheral heavy ion collisions as a photonuclear collider
and study vector meson production in this process. Moreover, such processes  can be also studied outside the heavy ion mode. For instance,  it has been discussed  in Ref. \cite{Klein:2003vd} the photoproduction of heavy vector mesons in $p\bar{p}$ collisions at the Fermilab Tevatron and in the $pp$ collisions at CERN LHC, since energetic protons also have large electromagnetic fields. These photoproduction reactions probe the gluon distribution in the proton at very small-$x$ values \cite{Goncalves:2001vs}, which open a new window to study parton saturation effects in exclusive processes.  In a separated
publication we will study these possibilities, considering the 
approaches discussed in this work.

\section*{Acknowledgments}
 The authors are grateful to Ruben Sandapen (D\'epartement de Physique et d'Astronomie, Universit\'e de Moncton) for his invaluable help and comments. Special thanks go  to Michael Lublinsky (DESY Theory Group, DESY) and Eran Naftali (Tel Aviv University, Tel Aviv) for providing us with the numerical results of the GLLMN model for the $J/\Psi$ photoproduction cross section. Useful discussions and valuable help of Igor Ivanov (Sobolev Institute of Mathematics, Novosibirsk State Un.)  at early stages of this work are gratefully acknowledged. The authors also thank Prof. Nikolai Nikolaev (IKP, Forshcungszentrum Juelich and L.D. Landau Institute) for calling our attention for the pioneering papers on vector meson production within the color dipole approach and their further developments. One of us (M.V.T.M.) thanks the support of the High  Energy Physics Phenomenology Group, GFPAE IF-UFRGS, Brazil. This work was partially financed by the Brazilian funding agencies CNPq and FAPERGS.

\end{document}